\documentclass[aps,prd,twocolumn,superscriptaddress,nofootinbib,preprintnumbers]{revtex4}

\usepackage{epsfig}
\usepackage{amssymb}
\usepackage{amsmath}
\usepackage{amsfonts}
\usepackage{hyperref}

\newcommand{\be}{\begin{equation}}
\newcommand{\ee}{\end{equation}}
\newcommand{\bea}{\begin{eqnarray}}
\newcommand{\eea}{\end{eqnarray}}
\newcommand{\bi}{\begin{itemize}}
\newcommand{\ei}{\end{itemize}}
\newcommand{\ben}{\begin{enumerate}}
\newcommand{\een}{\end{enumerate}}

\def\gsim{\mathrel{\rlap{\lower4pt\hbox{\hskip1pt$\sim$}}
    \raise1pt\hbox{$>$}}}         
\def\lsim{\mathrel{\rlap{\lower4pt\hbox{\hskip1pt$\sim$}}
    \raise1pt\hbox{$<$}}}         

\begin{document}

\preprint{TTP19-037}

\title{On dispersion relations and  hadronic light-by-light scattering
  contribution\\ to the muon anomalous magnetic moment}

\author{Kirill Melnikov}
\email{kirill.melnikov@kit.edu}
\affiliation{Institute for Theoretical Particle Physics, Karlsruhe Institute of Technology, Karlsruhe, Germany}
\author{Arkady Vainshtein}
\email{vainshte@umn.edu}
\affiliation{FTPI and School of Physics and Astronomy, University of Minnesota, Minneapolis, USA }
 \affiliation{KITP, UCSB, Santa Barbara, USA}

\begin{abstract}
  We discuss the use of dispersion relations for the evaluation of the pseudoscalar contributions to the
  muon anomalous magnetic moment. We point out that, in the absence of experimental data, reconstruction
  of light-by-light scattering amplitudes from their
  absorptive parts  is ambiguous and requires additional theoretical input. The need for an additional  input
  makes dispersive computations of the hadronic light-by-light scattering
  contribution to $g-2$ akin to phenomenological models, in spite of pretense  to the contrary. 
  In particular, we argue that the  recent proposal \cite{them}, based on the dispersive approach,
  satisfies  short distance constraints at the expense of  unjustifiably large deviations from the 
  chiral limit. 
  \end{abstract}

\maketitle


The measured value of the muon anomalous magnetic moment $a_\mu$ \cite{ben}
disagrees with the theoretical prediction for this
quantity, computed within the
Standard Model, by slightly more than three standard deviations or \mbox{${\cal O}(240)\! \times\! 10^{-11}$}.
If new experiment \cite{newexp}, currently underway at FNAL,  will confirm the results of the previous measurements,
a smaller error of the FNAL measurement will increase the discrepancy to about $5~\sigma$. To claim  
an even  larger significance
of the deviation will require either a shift in the measured value of $a_\mu$ or 
a reduction of the   theory error that, currently, is close to $50 \times 10^{-11}$. 

Moreover, since significant contribution to the  muon anomalous magnetic moment 
arises from kinematic regions where low-energy hadronic interactions
are important, a continuous scrutiny of theoretical assumptions,  which are behind the predicted value  of $g-2$,  is essential.
Among the various Standard Model  contributions to $g-2$, the so-called hadronic light-by-light scattering contribution is the
one that   is the most worrisome.  This is so because 
experimental information about hadronic contributions to the 
Green's function of three electromagnetic currents in the background of a soft magnetic field
\be
\begin{split} 
&(2\pi)^{4}\delta^{4}(q_{1}\!+\!q_{2}\!+\!q_{3}\!-\!k)\;T^{\mu \nu \alpha}(q_1,q_2,q_3)
\!=\!\! -\!\!\int\! \! {\rm d}^4 x\,{\rm d}^4 y\, {\rm d}^4 z \\
&\times e^{-i(q_1 x +q_2 y + q_3 z)}\big\langle 0 | T j_{\rm em}^{\mu}(x) j_{\rm em}^\nu(y)  j_{\rm em}^\alpha(z) |\gamma(k,\epsilon) \big \rangle, 
\label{eq1a}
\end{split} 
\ee
is, at best, very limited. 
In Eq.(\ref{eq1a}), the soft magnetic field is modeled by a transition from a vacuum to a soft photon
with momentum $k$ and polarization vector $\epsilon$; 
only linear terms in the  soft photon momentum $k$  must be  retained in Eq.(\ref{eq1a}). Also, we note that  throughout
the paper we consider all momenta to be outgoing. 

Original computations of the hadronic light-by-light tensor Eq.(\ref{eq1a})
and its contributions to the muon anomalous magnetic
moment were performed \cite{original} using models
where interactions of photons with photons were  described  by exchanges of relatively light  hadrons;  such features of 
QCD as the large-$N_c$ counting \cite{thooft}  and the chiral limit  were employed
for guidance \cite{nyf}.

Nevertheless, it felt very desirable to develop a way to evaluate  hadronic
light-by-light scattering contribution to the muon anomalous magnetic
moment that is based on first principles,    to reduce the model dependence as much as  possible. 
In (recent) years, two approaches to this problem emerged. One approach is based on the idea that the hadronic tensor
in Eq.(\ref{eq1a}) can be
unambiguously   predicted in certain  kinematic limits and that models for hadronic
light-by-light scattering contribution should conform to  these predictions.  A particularly  useful
application of this approach
in connection 
with the muon magnetic anomaly,
is the asymmetric limit $q_1^2 \sim q_2^2 \gg q_3^2$, where $q_3^2$ can be either larger than or comparable to
$\Lambda_{\rm QCD}^2$ \cite{us}. 
By combining these limits, we arrived at an unambiguous prediction for the
longitudinal contribution to the Green's function in the chiral limit. We have also used the perturbative regime 
 and the operator product expansion (OPE) to constrain the transversal  contribution as well. We employed
pion, $\rho$-, $\omega$- and $a_1$-mesons to construct a minimal model for the longitudinal and transversal structure functions
which  satisfies  short-distance  and non-perturbative constraints in the asymmetric kinematic limit \cite{us}. 

Another way to  compute the hadronic light-by-light scattering tensor in Eq.(\ref{eq1a}) from first principles
involves dispersion relations \cite{them,disp} that represent   $T^{\mu \nu \alpha}$ 
as  integrals of  its absorptive parts over certain kinematic variables. While formally exact, 
dispersion relations become useful in practice  if the absorptive parts of the corresponding Green's functions can be directly
obtained   from experimental data.  This is exactly what happens in a simpler case of the hadronic vacuum polarization
where data on $e^+e^-\!\to {\rm hadrons}$ annihilation cross section, in dependence on energy,  can be
used to determine the hadronic contribution to the Green's function
\be
\Pi^{\mu \nu}(q) = i\!\int\! {\rm d}^4x \,e^{-iqx} \langle 0 | T  j_{\rm em}^{\mu}(x) j_{\rm em}^\nu(0) | 0 \rangle, 
\ee
using dispersion relations.  The situation with the light-by-light tensor is obviously more complicated since
not only are the dispersion relations more complex, but also the absorptive parts are not quite known experimentally. 
Therefore, although 
dispersion representations of  Green's functions in general and in case of  the hadronic tensor in Eq.(\ref{eq1a})
in particular,  are undoubtfully correct,  it is not possible to use them in case of the
hadronic light-by-light scattering
contribution  without additional theoretical input required
to construct  absorptive parts.  Because of that, it is  our opinion, that the use
of dispersion relations {\it per se} does not add much to a better understanding of the hadronic light-by-light contribution
to $g-2$, compared to original computations \cite{original,us}.

It is instructive to compare the two approaches by studying contributions of pseudoscalar mesons to the hadronic
light-by-light scattering tensor Eq.(\ref{eq1a}). 
In the dispersive approach,\footnote{We restrict ourselves to the $\pi^{0}$ contribution; discussion
  of other pseudoscalars is largely identical.} where one looks for
a discontinuity with respect to a  variable $q_3^2$  \cite{them}, this contribution, effectively, arises from the
process $\gamma^*(q_1) \gamma^*(q_2) \to \pi^*_0(q_3) \to \gamma^*(q_3) \gamma(k)$. The result is proportional to 
\be
 F_\pi(q_1^2,q_2^2)\, \frac{1}{q_3^2 - m_\pi^2 }\, F_\pi(q_3^2,0),
\label{eq3}
 \ee
where the pion transition form factors refer to a process $\gamma^*(q_1) \gamma^*(q_2) \to \pi^0$. It is interesting to point
out that the  second pion  form factor  in Eq.(\ref{eq3}),  $F_\pi(q_3^2,0)$,
is conventionally interpreted as  a transition of
a photon with momentum $q_3^2$ and another on-shell photon to a pion on the mass shell. However, in case of $g-2$ the required
kinematics is different since the on-shell photon is, actually, {\it soft}. This implies that, for $q_3^2 \ne m_\pi^2$,
$F_\pi(q_3^2,0)$ can not be a regular pion transition form factor.  In line with this observation,
the dependence of
this  form factor on $q_3^2$ is ambiguous within  the dispersive approach since
\be
\begin{split} 
& F_\pi(q_1^2,q_2^2)\, \frac{1}{q_3^2 - m_\pi^2 }\, F_\pi(q_3^2,0) 
  = F_\pi(q_1^2,q_2^2) 
\\
& \times \Bigg [ \frac{F_\pi(m_\pi^2,0)}{q_3^2 - m_\pi^2 } 
    + 
  \frac{ F_\pi(q_3^2,0) - F_\pi(m_\pi^2,0)}{q_3^2 - m_\pi^2 } 
  \Bigg ].
\end{split} 
\label{eq4}
\ee
Clearly, the second term in square brackets
on the r.h.s. of Eq.(\ref{eq4}) is non-singular  at $q_3^2 = m_\pi^2$ and, therefore,  can not be associated
with the pion  pole, whereas  the first term does not have an ambiguous transition form factor anymore.  
The second term in Eq.(\ref{eq4}) can not be obtained from the pion pole discontinuity and requires 
 and requires additional information about Green's functions $\gamma^* \gamma^* \to \gamma^* \gamma$.  

It is said sometimes that the dispersion relation in Eq.(\ref{eq3})
cannot be derived   in the {\it reduced} kinematics,  when the photon $k$ 
is soft. Instead, Eq.(\ref{eq3}) should be obtained as  a $q_4 \to 0$ limit of a dispersion representation
of  the full $2 \to 2 $ process $\gamma^*(q_1) \gamma^*(q_2) \to \gamma^*(q_4) \gamma(q_3)$. 
A dispersive reconstruction of the $\pi_0$ pole contribution to  the full amplitude gives 
\be
F_\pi(q_1^2,q_2^2) \frac{1}{(q_3+q_4)^2 - m_\pi^2}F_\pi(q_3^2,q_4^2),
\ee
with obvious constraints on $q_{1,..,4}$ in the form factors  $(q_1 + q_2)^2 = m_\pi^2$ and $(q_3+q_4)^2 = m_\pi^2$. 
If we take the soft limit $q_4 \to 0$, the first constraint remains unaffected and covers large phase space of possible
values of $q_1$ and $q_2$, while the second one immediately implies $q_3^2 = m_\pi^2$, turning the  form factor $F_\pi(q_3^2,q_4^2)$
into a constant $F_\pi(m_\pi^2,0)$. Hence, independent of where one starts, the second form factor in Eq.(\ref{eq3})
does not follow from the dispersive  reconstruction of the $\pi_0$-pole contribution to the
hadronic light-by-light scattering
tensor in  the kinematics relevant for muon $g-2$. 

Additional information such as subtraction terms  that is  {\it naturally} 
missed by  the dispersive reconstruction of the pseudoscalar pole contribution can be obtained 
by studying the light-by-light tensor in a particular kinematic limit where  {\it  full non-perturbative
  description} of an important part of the hadronic tensor in the chiral limit can be achieved. 
Indeed, as explained in Ref.\cite{us}, it is beneficial
to study  an asymmetric kinematic limit $q_1^2 \approx  q_2^2 \gg q_3^2$. 
In that limit the OPE of the product of two electromagnetic currents is given by the axial-vector  current $j^{5}_{\alpha}$.
Then,  the light-by-light amplitude is reduced to a  triangle amplitude that describes
a transition of an axial-vector current to a soft photon and vector current, $j^{5}_{\rho}\to \gamma^{*}(q_{3})\gamma(k)$.
This transition amplitude can be decomposed into contributions
with definite SU(3) quantum numbers,  corresponding to isovector, octet and singlet axial-vector  currents. We write 
\be
T_{\mu \nu \alpha}(q_1,q_2,q_3) =
\frac{8}{\hat q^2} \epsilon_{\mu \nu \delta \rho} \hat q^\delta\!\!
\sum \limits_{a=3,8,0} W^{(a)} T^{(a)}_{\rho \alpha}(q),
\label{eq5a}
\ee
where  $\hat q=(q_{1}-q_{2})/2$, $q=q_{3}$ and
\be
\begin{split} 
T^{(a)}_{\alpha  \mu}  =
&
\,w^{(a)}_L(q^2) q_{\alpha} q^{\sigma} \tilde f_{\sigma \mu}
  \\
& + w^{(a)}_T(q^2) \left ( q^2 \tilde f_{\alpha \mu} - q_\mu q^{\sigma}\tilde f_{\alpha \sigma} - q_\alpha q^{\sigma} \tilde f_{\sigma \mu} \right ).
\end{split} 
\label{eq6a}
\ee
In Eq.(\ref{eq6a}) we  used
$\tilde f_{\alpha \beta} =  1/2 \epsilon_{\alpha \beta}  f^{\alpha \beta}$
and $f_{\alpha \beta} = \epsilon_\alpha k_\beta - k_\alpha \epsilon_\beta$.  

The amplitude $T_{\alpha \mu}$ at this point is defined {\it non-perturbatively}
as a matrix element of the time-ordered product of the axial and a vector current between the vacuum and the soft photon 
\be
T^{(a)}_{\alpha \mu}(q) = \!\int \!{\rm d}^4 x e^{i qx } \big\langle 0 | T j^{5(a)}_\alpha(x) j^{\rm em}_\mu(0) | \gamma(k,\epsilon)  \big\rangle.
\label{eq4}
\ee
By matching Eq.(\ref{eq5a}) to the short-distance limit of the light-by-light scattering amplitude 
one finds 
\be
w^{(3)}_{\rm L }(q^2) = 2 w^{(3)}_{T}(q^2),\;\;\;\; w^{(3)}_{L}(q^2) = -\frac{2}{q^2},
\ee
for the isovector contribution. 
Perturbatively, in the chiral limit there are no corrections to the longitudinal structure function
$w^{(3)}_{L}$, thanks  to the Adler-Bardeen theorem \cite{ab}, 
and to the  transversal one $w^{(3)}_{T}$\!, due a peculiar relation between $w^{(3)}_{T}$ and
$w^{(3)}_{L}$ in the chiral limit discovered  in \cite{va}.

Furthermore, it is possible to compute  non-perturbative corrections to the Green's function in Eq.(\ref{eq4}) by 
performing an operator product expansion. It follows from this analysis \cite{us},
that the longitudinal form factor $w_L$ {\it does not receive any corrections
  in the chiral limit} whereas there are non-perturbative
corrections to the transversal form factor.  This information  has been
used to construct models for $w_L$ and $w_T$ using small number of
light  mesons that contribute in a pseudoscalar and vector channels.   The form factors read \cite{us}
\be
\begin{split} 
&  w^{(3)}_{L}(q^2) = -\frac{2}{q^2 - m_\pi^2},\;\;\;\;\;
  \\
& w^{(3)}_T(q^2) = \frac{1}{m_{a_{1}}^2 - m_{\rho}^2} \left [ \frac{m_\rho^2}{q^2 - m_\rho^2}
  - \frac{m_{a_{1}}^2}{q^2 - m_{a_{1}}^2} \right ].
\label{eq6}
\end{split} 
\ee
The mass of the pion is added to $w^{(3)}_L$ to go beyond the chiral limit at small $q\sim m_{\pi}$. One may be concerned 
that this step does not properly account for possible chiral-violating  effects in $w^{(3)}_L$ at  $q^2\gg m^2_{\pi}$. Such concerns
are, however, unfounded; see the discussion  at the end of this paper. 

The model Eq.(\ref{eq6}) was recently criticized in Ref.~\cite{them} on the basis that it ``distorts'' the low-energy
``dispersive'' formula shown in Eq.(\ref{eq3}) where the ``distortion'' means the absence of the second
form factor $F_\pi(q_3^2,0)$ in the ``right'' version of Eq.(\ref{eq3}). 
It was  suggested in Ref.~\cite{them} that
one can ``repair'' the model of Ref.~\cite{us} by allowing for an infinitely large number of excited pions to contribute to
the light-by-light scattering amplitude. The constraint  $w_L(q^2) = -2/q^2$ at $q^2 \gg \Lambda_{\rm QCD}^2$ is then enforced
by a particular choice of form factors. 

In what follows we would like to explain why the two claims made in Ref.~\cite{them}, i.e.   the incompatibility
of Eq.(\ref{eq6}) with dispersion relations and the possibility to ``correct'' this incompatibility by allowing infinite number
of exchanges in the longitudinal form factor,  are  strongly questionable.

To address the problem of incompatibility with dispersive reconstruction
, we note that, as already pointed out after Eq.(\ref{eq4}) the dispersion
relations in $q_3^2$ are ambiguous and the presence or absence  of the  form factor $F_\pi(q_3^2,0)$ is a question
about non-pole  rather than an  absorptive or pole  part.  These additional terms do not follow from dispersion
relations applied to the pion pole and require additional information. Given this ambiguity, it is impossible to 
discuss the issue of an incompatibility seriously. 

Moreover, in the asymmetric kinematics
$q_1^2 \sim  q_2^2 \gg q_3^2$, the dispersion reconstruction in the variable $q_3^2$ should be  done for a
transition amplitude shown in  Eq.(\ref{eq4}) which, essentially, describes a mixing of vector and vector-axial
currents in a constant  magnetic field. To provide a dispersion
reconstruction of this Green's function we simply saturate it in the spirit of the vector dominance by allowing
vector and axial currents to ``mix'' into lightest mesons with relevant  quantum
numbers.  These mesons then have {\it point-like} interactions with
the soft photon and the remaining current. 

The lightest mesons that we consider are  the pion and the $a_1$ axial-vector meson,
that can mix into the axial current, and the $\rho,\,\omega$ mesons that can mix into the vector current.
Ignoring the   (tiny) mass difference between
$\rho$ and $\omega$ mesons, we can write  their contribution to the Green's function as 
\be
\begin{split} 
  & T^{(3)(\rho)}_{\alpha \mu}(q^2) = c_L^{(\rho)} q_{\alpha} q^{\sigma} \tilde f_{\sigma \nu} \frac{-g^{\nu}_{\mu}}{q^2 - m_\rho^2}
  \\
  & + c_T^{(\rho)}  \left ( q^2 \tilde f_{\alpha \nu} - q_\nu q^{\sigma} \tilde f_{\alpha \sigma} - q_\alpha q^{\sigma} \tilde  f_{\sigma \nu} \right )
  \frac{-g^{\nu}_{\mu}}{q^2 - m_\rho^2},
  \label{eq10}
\end{split} 
    \ee
    where $c_{L,T}^{(\rho)}$ are unknown constants. Note, that we discarded $q_\mu q_\nu/m_\rho^2$ term in the numerator
    of the $\rho$ propagator in Eq.(\ref{eq10}),  
    since  the amplitude Eq.(\ref{eq4}) is constructed in such a way that the vector current is conserved. 

A similar contribution of an $a_1$ axial-vector meson reads
\be
\begin{split} 
  & T^{(3)(a_{1)}}_{\alpha \mu}(q^2) = c_L^{(a_{1)}} \frac{-g_{\alpha}^{\beta} + ({q_\alpha q^{\beta}/m_{a_1}^2}) }{q^2 - m_{a_1}^2}
   \,q_{\beta} q^{\sigma} \tilde f_{\sigma \mu} 
  \\
  & + c_T^{(a_{1})} \frac{-g_{\alpha}^{\beta}}{q^2 - m_{a_1}^2}
  \left ( q^2 \tilde f_{\beta \mu} - q_\mu q^{\sigma} f_{\beta \sigma} - q_\beta q^{\sigma} f_{\sigma \mu} \right )
    \end{split} 
\label{eq11}
\ee
    Note that in this case $q_\alpha q_\beta/m_{a_1}^2$ can be discarded for the transversal part but it plays an important
    role in  the longitudinal part. Indeed, we obtain
    \be
    \frac{-g_{\alpha \beta} + (q_\alpha q_\beta/m_{a_1}^2) }{q^2 - m_{a_1}^2} \; q^{\beta} =
    \frac{q_\alpha}{m_{a_1}^2}, 
    \ee
    which implies that the pseudo-vector meson does not contribute to the longitudinal part of the Green's function {\it
      in a dispersive sense} since the vector-axial  pole disappeared. 

    Adding a pion-pole  contribution to the longitudinal form factor leads to the following result for  the
    dispersively  reconstructed   structure function 
    \be
\begin{split} 
  & w^{(3)}_L(q^2 ) = -\frac{2}{q^2} - \frac{c_L^{(\rho)}}{q^2 - m_\rho^2 },
  \\
  & w^{(3)}_T(q^2) = -\frac{c_T^{(\rho)}}{q^2 - m_\rho^2 } - \frac{c_T^{(a_{1})}}{q^2 - m_{a_1}^2}.
  \label{eq13}
\end{split} 
  \ee
  At this point, the constants $c_{L,T}^{(\rho,a_{1})}$ are arbitrary. For example, by choosing $c_L^{(\rho)} = -2$, 
we obtain 
  \be
w_L(q^2) = -\frac{2}{q^2} \;\frac{-m_\rho^2}{q^2-m_\rho^2}.
  \ee
  Hence,  this choice of an ``effective''
  coupling constant introduces a ``pion form factor''  into a longitudinal structure
  function $w_L(q^2)$. Such a choice will be in accord with Eq.(\ref{eq3}) and would imply a stronger
  suppression of the longitudinal contribution for $q_3^2 \gg m_\rho^2$ than what is allowed by  perturbative and
  non-perturbative    matching. 

  However, it is quite obvious from Eq.(\ref{eq10}) that other choices are possible and are, in fact, better
  motivated. As we proved in Ref.~\cite{us},  the longitudinal structure function $w_L(q^2)=-2/q^2$ is  {\it exact} in the
  chiral limit and it is not renormalized by either perturbative or non-perturbative corrections. Clearly, the choice
  $c_L^{(\rho)} = -2$ leads to $w_L(q^2)$ that violates this assertion.  To comply with it, we have chosen $c_L^{(\rho)} = 0$.
   It is also seen from Eq.(\ref{eq10}) that the function $w_T(q^2)$ can be reconstructed independently
  of the longitudinal one. The coefficients $c_T^{(\rho,a_{1})}$ in those cases are fixed  by requiring that at large
  values of $q^2$ the transversal function
  $w_T$ matches perturbative asymptotic and that non-perturbative corrections at large $q^2$ are
  consistent with the operator product expansion \cite{us}.

  One may be wondering if models that include more resonances can satisfy the asymptotic behavior required by short-distance
  QCD \cite{them}. To analyze this question in a simple setting, 
  suppose that we include yet another $\rho$-meson into a dispersive reconstruction of the longitudinal
  function $w^{(3)}_L$. We find 
  \be
  w^{(3)}_L(q^2 ) = -\frac{2}{q^2} - \frac{c_L^{(\rho)}}{q^2 - m_\rho^2 } - \frac{c_L^{(\rho_1)}}{q^2 - m_{\rho_1}^2 }.
    \ee
    The large-$q^2$ asymptotic of the short-distance constraint requires $c_L^{\rho} = -c_L^{\rho_1}$. Then $w^{(3)}_L$ becomes
    \be
    w^{(3)}_L(q^2 ) = -\frac{2}{q^2} - c_L^{\rho} \frac{m_\rho^2 - m_{\rho_1}^2}{(q^2 - m_\rho^2) (q^2 - m_{\rho_1}^2)},
    \label{eq17}
    \ee
    and $c^\rho_L$ remains unconstrained. To constrain it,  we note
    that the longitudinal structure function should be equal to $w^{(3)}_L = -2/q^2$ {\it non-perturbatively} in the chiral
    limit. Hence, we obtain the constraint
    \be
   \lim_{m_{u,d,s} \to 0} c_L^{\rho} = 0.    
    \ee
    We note that the model of Ref.\cite{them} {\it violates} the above equation and   claims, effectively, that
    $c_L^{\rho} \sim 1$ also 
    in  the chiral limit. 
    
  In  order to  account for the violation of chiral symmetry at $q \sim m_{\pi}$ we added a pion mass to
  the $1/q^{2}$ pole in $w^{(3)}_{L}$,  $1/q^2 \to 1/(q^2 - m_\pi^2)$, see Eq.(\ref{eq6}). This modification introduces
  a deviation from the  asymptotic behavior of $w_L$  and implies
  \be
 c_{L}^{\rho} \sim \frac{m_\pi^2}{m_\rho^2}.
  \ee
   One may wonder if this estimate  is reasonable.  We addressed  this question in  Ref.~\cite{book}
  using the OPE for the longitudinal structure functions. It was shown there that non-perturbative  corrections
  to $w^{(3)}_L$ are generated by an operator ${\cal O}_{\alpha \beta} = -i \bar q \sigma_{\alpha \beta} \gamma_5 q$
  and are proportional to quark masses. 
  The matrix element of the  operator ${\cal O}_{\alpha \beta}$ between the photon and the vacuum can
  be estimated by expressing it  through the magnetic susceptibility of the quark condensate.
  Numerically, it leads to the   change  $\Delta w^{(3)}_{L}$ in the longitudinal $w^{(3)}_{L}$,
  \be
  \frac{\Delta w^{(3)}_{L}}{w^{(3)}_{L}}=\frac{(0.18\; {\rm GeV})^{2}}{q^{2}}\,,
  \ee
  which supports introduction of the pion mass as the {\it only} source of the chiral symmetry
  violation and   does not leave any  room for the
  proposal of  Ref.~\cite{them}.
  
  It is clear that  our arguments concern a special kinematic region $q_1^2 \sim q_2^2 \gg q_3^2$ and one may ask  if this
  kinematic region is sufficient for the evaluation of the hadronic light-by-light scattering contribution to $g-2$ with
  sufficient accuracy. 
  However, there is no doubt that (a) this region provides the largest contribution to $a_\mu^{\rm hlbl}$; 
  (b) it  allows for an {\it exact}  non-perturbative analysis of the longitudinal structure
  function in the chiral limit and (c) it supplies strong evidence that corrections
  to the chiral limit are {\it small}.
  It is our view that the above points provide a  motivation for using the asymmetric kinematic limit
  as a  diagnostic tool to  check the validity of  different models. Unfortunately, the model of Ref.~\cite{them}, whose authors
  purport \cite{them} to remove the ``biggest systematic uncertainty due to short-distance constraints''
  with the result ``that the asymptotic
part of the hadronic light-by-light  tensor is under sufficient control for
the first release from the Fermilab experiment,''    fails to pass the test. 
  
{\bf Acknowledgments}
We thank G.\,Colangelo and M.\,Einhorn for useful discussions.
A.V. is grateful to Kavli Institute for Theoretical Physics at UC Santa Barbara for their support and hospitality.
His research was  supported in part by the National Science Foundations under Grant No. NSF PHY-1748958.


\end{document}